\documentclass{llncs}
\usepackage{amsmath}
\usepackage{amssymb}
\usepackage{amsfonts}
\usepackage{amstext}
\usepackage{latexsym}
\usepackage{graphicx}

\newtheorem{propn}{Proposition}{}{}
\newtheorem{thm}{Theorem}{}{}
{}{}
\newtheorem{cor}{Corollary}{}{}
{}{}
\newtheorem{Def}{Definition}{}{}
{}{} 
{}{}
{}{}

\newcommand{\commentout}[1]{}

\newcommand{\kleene}[1]{{#1}^{*}}
\newcommand{\union}{\cup}
\newcommand{\set}[1] {\{#1 \}}

\newcommand{\inv}[1]{#1^{-1}}

\newcommand{\bcolvec}{\left(\begin{array}{c}}
\newcommand{\ecolvec}{\end{array}\right)}
\newcommand{\bmat}[1]{\left(\begin{array}{#1}}
\newcommand{\emat}{\end{array}\right)}

\newcommand{\imp}{\Rightarrow}

\newcommand{\be}{\begin{enumerate}}
\newcommand{\ee}{\end{enumerate}}
\newcommand{\beq}{\begin{equation}}
\newcommand{\eeq}{\end{equation}}
\newcommand{\beqx}{\begin{displaymath}}
\newcommand{\eeqx}{\end{displaymath}}
\newcommand{\beqa}{\begin{eqnarray}}
\newcommand{\eeqa}{\end{eqnarray}}
\newcommand{\beqax}{\begin{eqnarray*}}
\newcommand{\eeqax}{\end{eqnarray*}}

\newcommand{\cali}[1]{{\mathcal #1}}

\begin{document}

\title{An Algebraic Characterization of Security of Cryptographic Protocols}
\author{Manas K. Patra and Yan Zhang}
\institute{
Department of Computing and Mathematics, University of Western Sydney, \\ 
Locked Bag 1797, Penrith South DC, NSW 1797 \\
Australia \\ 
}
\date{}
\maketitle 

\begin{abstract}
Several of the basic cryptographic constructs have associated algebraic structures. Formal models proposed by Dolev and Yao to study the (unconditional) security of public key protocols form a group. The security of some types of protocols can be neatly formulated in this algebraic setting. We investigate classes of two-party protocols. We then consider extension of the formal algebraic framework to private-key protocols. We also discuss concrete realization of the formal models. In this case, we propose a definition in terms of pseudo-free groups.  

\noindent
{\em Keywords/Topics}: security, public key cryptosystem, free and pseudo-free groups and monoids.
\end{abstract}

\section{Introduction and Background} 
The present paper explores some algebraic structures inherent in several classes of security protocols. Such structures have been known to exist. For example, the set of possible messages over some alphabet $\cali{A}$ constitute a {\em free monoid} $\cali{A}^*$. The encryption and decryption operations must be inverse of each other. If we consider them as mappings $\cali{A}^* \rightarrow \cali{A}^*$ they form a group. Moreover, many encryption schemes are based on some well-known algebraic structures. The RSA encryption is a bijective map $Z_n \rightarrow Z_n$, where $Z_n$ is the {\em ring} of integers modulo $n$. So we have, on the one hand, formal models of classes of protocols which carry algebraic structures and on the other concrete realizations of these models based on some sets with inherent algebraic structures. One of the basic issues addressed in this paper is the notion of {\em security} of protocols in the algebraic setting. In the formal model where we assume perfect encryption the security is unconditional. Hence, it can be breached due to a faulty design of protocols. In the concrete model however the encryption is based on the assumption that certain tasks are computationally infeasible. In this case, the security can be compromised due to a faulty design {\em or} some hidden relations among the basic operators. Although protocols based on PKC are believed to be secure against passive attacks, an improperly designed protocols may be compromised by an active adversary, as first pointed out by Needham and Schroeder \cite{NS}. The analysis of all possible such attacks requires some level of abstraction and formalization. Such a formalization was first given in the seminal work of Dolev and Yao \cite{DY} (referred to as DY).  The class of protocols discussed in DY are two-party cascade protocols in which two users exchange messages back and forth. Other notable early works dealing with the formal approach to security include \cite{Ban,Lowe}. 

An alternative approach to security is the computational approach. Informally, a protocol is considered secure if it is computationally infeasible for the adversary to acquire any useful information \cite{Goldreich,AbadiRog}. The computational approach is more difficult in proving security of protocols. Starting with the work of \cite{AbadiRog} there has been extensive work to relate the two views of cryptography. In  \cite{AbadiRog} the authors first give the formal framework for some cryptographic primitives. In terms of security, their main result roughly translates to the following: if we can show that a protocol, formally an expression, is equivalent to another expression over a fixed string then the protocol is secure since it is infeasible for the adversary  to distinguish between the actual plaintext and an arbitrary bit string. Thus a formal system is sound if formal indistinguishability ({\bf FI}) implies computational indistinguishability ({\bf CI}). The converse ({\bf CI} $\imp$ {\bf FI}) is called the {\em completeness} of the formal system. It has been proved for the Abadi-Rogaway formal system under some extra assumptions \cite{WarMicc1}. The works \cite{AbadiRog,WarMicc1} dealt with symmetric (private) key encryptions and passive adversaries and in \cite{WarMicc2} the authors prove soundness of a formal system similar to \cite{AbadiRog} for public key cryptosystem with {\em active} adversaries. The work \cite{WarMicc2} deals with issues that are closest to the current work. 

In this work we take a fresh look at the DY model. We investigate algebraic structures associated with a class of protocols based on public key cryptosystems. We observe that the model defined by strings of operators can be given the structure of {\em group} called the Dolev-Yao (DY) group. The main results of this paper are characterization of the security of the protocols in these group structures. Specifically, we show that a set of elements (strings) defining the protocol is insecure if and only if they contain a subgroup. This is strictly true in the abstract setting when we assume that there are no special relations among the elements- the DY group is {\em free}. In a concrete realization there will be some relations among the group elements. We propose extensions of the notion of security in terms of {\em pseudo-free} groups rather than free groups. We also consider extension to private key cryptosystems. 

We first review the Dolev-Yao model. One defines the abstract setting of a protocol in terms of some basic operations (encryption, decryption, nonces etc.). These operations form a monoid. Then a protocol is simply a sequence of words, the elements of the monoid. The security of a protocol is defined in terms of these words. Specifically, we show that a set of elements (strings) defining the protocol is insecure if and only if they contain a subgroup. We consider first the simple cascade protocols where the message texts are encrypted and decrypted straight without further operations like nonces. In this case the monoid turns out to be a group and a protocol is insecure if and only the elements defining it form a subgroup. Next, we consider protocols with nonces (name-stamps, date-stamps etc.). The algebraic characterization is trickier here because some of the operations are undefined in a real implementation. We show that even in this case we can sensibly define a monoid of operations and characterize protocol security in terms of some algebraic condition. We use the algebraic characterizations to prove some general theorems on secure and insecure protocols. We apply these results to some well-known protocols. We also discuss the concrete realization of the cryptosystems.  We analyze the implication on security in this situation. The problem of security in an arbitrary realization is undecidable since it can be reduced to the {\em word problem} \cite{Rotman}. The final section discusses possible extensions of the definitions and methods. 
\section{The Dolev-Yao model}
In this section we review the essentials of the model proposed by Dolev and Yao. The first assumption is that we do not concern ourselves with the details of the public key cryptographic system. Further, we assume that we have a finite set of symbols $\cali{E}=\{E_1, E_2, \dotsc, E_n\}$ where $n$ is an integer. Informally, $n$ denotes the number of users in the network and $E_i$ represents the {\em public} encryption function of the $i^{\text{th}}$ user. Similarly we have another set $\cali{D}= \{D_1, \dotsc, D_n\}$ representing the {\em private} decryption function of the users. For example, if $K_i$ and $K'_i$ are the public and private keys of user $i$ then $E_i(M)= E(M,K_i)$ and $D_i(M)= D(M,K'_i)$, where $E$ and $D$ are the respective encryption and decryption functions and $M$ is the message text. We also add another operator, $I$ the ``identity'' operator. In general the encryption and decryption schemes need not be same for all users but they must satisfy $E_iD_i=D_iE_i=I$. We simply treat them as letters from some alphabet. For each pair of users $(i,j)$ define the sets \(A_{ij}= \{E_i,E_j,D_i\} \)
Informally, $A_{ij}$ represents the set of operators available to user $i$ in a two-party exchange between itself and user $j$. A two party cascade protocol is finite sequence of strings $\{\alpha_1,\alpha_2, \dotsc, \alpha_r\}$ and $\{\beta_1,\beta_2, \dotsc, \beta_{r'}\}$ where \(\alpha_i \in A_{ij}^* \text{ and } \beta_i\in A_{ji}^* ,\;1\leq i,j \leq n\) and $r'=r-1 \text{ or }r$. Intuitively, users $i$ and $j$ can use any number of layers of encryption and decryption and thus the set of operations available are included in $\cali{E}\union\cali{D}$. The definition of cascade protocols is a consequence of the following assumption on the protocols \cite{DY}.
\be
\item
It is a perfect public key system. Hence: 1.\ the functions $E_i$ are strictly one way: they are unbreakable, 2.\ the public directory is secure: the $E_i$ are fixed once for all, 3.\ everyone has access to all the encryption functions $E_i$, 4.\ only user $i$ knows $D_i$. 
\item
In the two-party protocol only the two parties concerned are involved in the communication; the assistance of a third party is not needed. 
\item
The protocols are uniform, that is, the same format is used by any pair of legitimate users. 
\item
Next we model the behavior of the adversary. We assume that the adversary is capable of active attacks. Specifically: 1.\ the adversary can intercept any message passing through the communication channels;2.\ he is a legitimate user and thus can initiate a dialog with other users; 3.\ he can successfully impersonate another user when necessary. 
\ee
We assume that the above assumptions are valid for {\em any} protocol (not just cascade protocols) unless stated otherwise. 

Next we describe the formal model for the protocols. Let $x,y$ be variables ranging through the set $J_n\equiv \{0,1,\dotsc, n\}$. A two-party cascade protocol is given by a pair of sequences
\begin{gather}
\{\alpha_1(x,y), \alpha_2(x,y),    \alpha_r(x,y)\} \text{ and } \{\beta_1(x,y), \beta_2(x,y),  \beta_{r'}(x,y)\}\\
\alpha_i(x,y) \in A_{xy} \text{ and } \beta_i(x,y) \in A_{yx} 
\end{gather} 
Further, define the sequences 
\beq
\begin{split}
N_1(x,y) = \alpha_1(x,y) &\quad N_2(x,y) = \beta_1(x,y)\alpha_1(x,y) \\
N_{2k-1}(x,y) = \alpha_{k}(x,y)N_{2k-2}(x,y) &\quad N_{2k}(x,y) = \beta_kN_{2k-1}(x,y) \\
\end{split}
\eeq
The intuition behind this abstract definition is the following. User $x$ initiates the dialog with $y$ by applying $\alpha_1(x,y)$ to the message $M\in \{0,1\}^{*}$. Then, $y$ responds with the application of $\beta_1(x,y)$, $x$ follows with $\alpha_2(x,y)$ and so on. In round $k$ ($k\geq 1$) user $x$ sends the message $N_{2k-1}M$ and in turn, receives the message $N_{2k}M$. For example, in the simple protocol  discussed later we have \(\alpha_1(1,2) = E_2, \text{ and } \beta_1(1,2)= E_1D_2\)
\newcommand{\prot}{\cali{P}}
Let $\prot$ be a two-party cascade protocol. Let $s$ be any user name (the adversary) and
\beq
\begin{split}
\Gamma_1 (s) = \cali{E}\union \set{D_s}, \; \Gamma_2 &= \{ \alpha_i(x,y)| \text{ for all } x\neq y \text { and } i \geq 2\} \text{ and } \\
\Gamma_3 &= \{ \beta_i(x,y)| \text{ for all } x\neq y \text { and } i \geq 1\} \\
\end{split}
\eeq
Next we define the security of a protocol. 
\begin{Def} \label{def:insecure}
A protocol $\prot$ is {\em insecure} if there is some string $\lambda \in \Gamma_1(s)\union\Gamma_2 \union \Gamma_3$ such that $\lambda N_k(i,j)= \epsilon $ for some $k$ ($\epsilon $ denotes the empty string).  
\end{Def}
See \cite{DY} for the motivation for this definition is as follows. If the protocol is insecure then the secret message can eventually be obtained by the adversary. 

\section{The Dolev-Yao group for cascade protocols}
We start this section with some standard algebraic definitions \cite{Rotman}. A semigroup is set $S$ with a binary operation or product $\circ$ that is associative ($a\circ(b\circ c) = (a\circ b)$). A monoid is a semigroup $\{S,\circ\}$ with an identity element $e$ ($e\circ a=a\circ e=a$). A group is a monoid $M$ such that every $a\in M$ has an inverse $a^{-1}$ ($a\circ\inv{a}=\inv{a} \circ a =e$). Below we suppress the symbol $\circ$ for the product. 
We have seen above that for cascade protocols the available operators are from $\cali{E}\union \cali{D}$. The set $\kleene{\cali{E}}$ (the Kleene closure of $\cali{E}$) is the set of words, including the empty word, formed by the alphabet $\cali{E}$. Now consider the {\em free group} generated by the set $\cali{E}$ \cite{MKS}. We recall the free group construction. Let $A$ be a set (the alphabet). Let $\inv{A}$ be another set, disjoint from $A$ such that there is a bijective correspondence $a \leftrightarrow a$ between the two. We write $\inv{A}= \{\inv{a}| a\in A\}$. Let $\epsilon$ be the empty string. Then we define a product on the set $S_A \equiv \kleene{(A\union \inv{A})}$ by concatenation ($\sigma\cdot \mu= \sigma\mu$) along with the relations $a\inv{a}=\inv{a}a=\epsilon$. That is, we replace $a\inv{a}$ and $\inv{a}a$ by $\epsilon$ in any string. More formally,  define an equivalence relation $\sim$ between two strings $\sigma$ and $\mu$ as: $\sigma\sim \mu$ if $\mu$ can be obtained from $\sigma$ by insertion or deletion of strings of the form $a\inv{a},\inv{a}a \text{ and }\epsilon$. Then the set $F(A)=S_A/\sim$, the set of equivalence classes is a group. For details see \cite{MKS}. For convenience, we continue to write the members of $F(A)$ as elements of $S_A$ rather than the equivalence class. For a free monoid we have only the set $A$ and the relation $\epsilon$. The essential property of a free group or monoid $F(A)$ over the set $A$ is that any mapping of the set $A$ into a group $G$ can be uniquely extended to a group homomorphism (see \cite{MKS} for details). Recall that a homomorphism between two monoids is a mapping that preserves the identity and products. A homomorphism between two groups is a homomorphism of the underlying monoids that preserves inverses. A submonoid $A$ of a monoid $M$ is a subset with identity that is closed under products. We call $F(\cali{E})$ the DY group. Further, we use $D_i$ and $\inv{E_i}$ interchangeably. 
A concrete realization of the DY group is given by the action of encryption and decryption operators on $\{0,1\}^{*}$, the set of binary strings. Thus, if $K_i, \text{ and } P_i$ are $i$'s public and private key respectively then $E_i(m)= E(m, K_i) \text{ and } D_i(m)=D(m, P_i)$. We note that a concrete realization of a free group may result in more relations. For example, for a commutative group we have the relations $ab\sim ba$. We further mention that a particular realization realization of the DY group in the RSA encryption scheme is {\em distinct} from the RSA group \cite{Rivest}. In general, the latter is commutative while the former is not. 

Let us consider an example discussed in \cite{DY}. User $i$ sends $j$ a message $m$ \((i,E_j(m), j)\) and then $j$ sends back the message $( j, E_i(m),i)$. This protocol is very easily broken. The adversary, henceforth denoted by $s$, intercepts the first message from $i$ and sends it to $j$. Then $j$ sends the message $( j, E_s(m),s)$. The adversary decrypts the message using $D_s$. It is easy to verify that in this case the the monoid generated sets $\Gamma_1= \set{D_s} \text{ and } \Gamma_2 = \set{E_sD_jE_j}$ a {\em subgroup} of DY. We will see that this is a general phenomenon for insecure protocols.  

\subsection{An algebraic characterization of security} 
In this section we come to the main theme of this work. Dolev and Yao gave a characterization of the secure cascade protocols in terms of properties of the strings $\alpha_i(x,y)$ and $\beta_j(x,y)$. We prove an equivalent characterization in the algebraic setting of the DY {\em group}. We can then deduce their characterization. In the following, the word generate will always imply the multiplicative set (a monoid). 
\begin{thm} \label{thm:alg_secure1}
Let $\prot$ be a two-party cascade protocol. Assume that the parties involved have names 1 and 2 and the adversary is $s$. Then, with the notation as above, $\prot$ is insecure if and only if there is a set $T\subseteq \set{E_1,E_2,E_s,D_s}\subset \Gamma_1(s)$ such  that one of the following condition holds. 
\be
\item
The set $\set{\alpha_1(x,y)} \union T$ generates a subgroup of DY multiplicatively.  
\item
The set $T\union \Gamma_2(x,y)\union \Gamma_3(x,y);x,y\in \set{1,2,s}$ generates a nontrivial subgroup of DY. 
\ee
where $\Gamma_j(x,y)$ denotes the set $\Gamma_2$ with specific users $x\text{ and }y$. 
\end{thm}
\begin{proof}
Let us first note that the first condition takes care of a rather trivial situation. It can only come about if the user $x$ initiates the conversation by sending the message {\em without} an encryption or if she applies her own decryption operator! In any case, it is clear that the protocol is insecure. Next, suppose the second condition holds. Then the set $T\union \Gamma_2(x,y)\union \Gamma_3(x,y)$ a subgroup $S$. In particular, $\inv{E_1},\inv{E_2} \in S$. Hence, there is a string $\lambda \in S$ such that $\lambda N_i=\epsilon$ since the latter is the identity element of the group. It follows from the definition \ref{def:insecure} that the protocol is insecure. This proves the sufficiency of the condition. 

To prove necessity of the condition assume that the protocol is insecure. Then there is some string $\lambda$ such that \(\lambda N_i= \epsilon, \;i\geq 1\)
First, suppose that $i=1$ and $N_1=\alpha_1$ does not contain $E_1\text{ or } E_2$. Then we must have $\alpha_1=\epsilon \text{ or } D_x^k$, for some integer $k$. In the first case, we obtain the trivial subgroup by choosing $T$ to be empty set and in the second case we choose $T=\{E_1\}$. In either case, the first condition of the theorem is satisfied. 

Now let $N_i,\;\;i\geq 1$ satisfy the above equation. Suppose $i=2j$ is even (the proof for the odd case is similar). Then 
\[
\begin{split}
N_{2j}(1,2) &=   (\beta_j(1,2)\alpha_j(1,2) \cdots \alpha_2(1,2)\beta_1(1,2))\alpha_1 (1,2)\\
&\equiv \phi_j (1,2)\alpha_1 (1,2)\text{ and } \\
\lambda &= \inv{\alpha_1}(1,2) \inv{\phi_j}(1,2)
\end{split}
\]
By assumption, $\lambda \in (\Gamma_1(s)\union \Gamma_2(x,y) \union \Gamma_3(x,y))^{*}\equiv H$. Let $H'=H\union \set{\alpha_1(1,2)}$. Clearly we may restrict to the set $\set{1,2,s}$ of users.  
Observe first that any $N_i(x,y)$ is of the form $E_x^{i_1}E_{y}^{j_1}E_x^{i_2}E_y^{j_2} \cdots E_x{i_m}E_{j_m}$ where $i_r \text{ and } j_r$ are integers. Recall that we identify $\inv{E_x}=D_x$. Suppose that all the exponents of  $E_1,\text{ and }E_2$ in the expansion of $N_{2j}(1,2)$ are non-negative. We may assume that at least one of them, say that of $E_1$, is positive (otherwise there is nothing to prove). Then by successive application of $E_1\text{ or }E_2$ we conclude that \(\inv{E_1} \text{ is in } H\). From the definition of the sets $\Gamma_2$ and $\Gamma_3$ we can interchange the role of $E_1$ and $E_2$ and we conclude that $\inv{E_2}$ is also a member of $H$. Choose $T=\set{ E_s, E_1, E_2, D_s}$. Then, $T\union \Gamma_2{1,2}\union \Gamma_3{1,2}$ generates a subgroup. Hence, we may assume that $N_{2j}(1,2)$ contains negative powers of $E_i,\;i=1,2$. In any case we have $N_{2j}(1,2)= \phi_j (1,2)\alpha_1 (1,2)$ and $\lambda= \inv{\alpha_1}(1,2) \inv{\phi_j}(1,2)$. As $\phi_j(1,2)\in H$ we conclude that $\inv{\alpha_1}\in H$.  Let 
\(\inv {\alpha_1} = E_1^{-i_1}E_{2}^{-j_1}E_1^{-i_2}E_2^{-j_2} \cdots E_1^{-i_m}E_2^{-j_m}\in H\) 
Where $i_k,j_k$ are integers. We recall that $\alpha_1$ may contain only $E_1,E_2,\text{ or }D_1$. Thus, no $j_k$ can be negative. We have assumed that not all of them are zero for otherwise we are back to the first condition. Therefore, we may write 
\(\inv{\alpha_1} = E_1^{i_1} D_2^{j_1} E_1^{i_2} D_2{j_2}\cdots E_1^{i_m}D_2^{j_m} \)
We assume that none of the exponents in the middle (that is, $j_1,i_2,\cdots,i_m$) are zero and consider several cases. 
As $\inv{\alpha_1}$ belongs to the set $H$, it must be of the form
\[ 
\begin{split}
\inv{\alpha_1} (1,2)& = {\boldsymbol \alpha}^{(a^1_1, \dotsc, a^k_1)}(1,2) {\boldsymbol \beta}^{(b^1_1, \dotsc, b^s_1)}(1,2)E_1^{c_1}E_2^{d_1} \\
&{\boldsymbol \alpha}^{(a^1_2, \dotsc, a^k_2)}(1,2) {\boldsymbol \beta}^{(b^1_2, \dotsc, b^s_2)}(1,2)E_1^{c_2}E_2^{d_2} \cdots \\
\end{split}
\]
where \({\boldsymbol \alpha}^{(a^1_1, \dotsc, a^k_1)}(1,2) \equiv \alpha_2^{a^1_1}(1,2)  \cdot \alpha_{k+1}^{a^k_1}(1,2)\) and  \({\boldsymbol \beta}^{(b^1_1, \dotsc, b^s_1)}(1,2) \equiv \linebreak \beta_1^{b^1_1}(1,2) \cdots \beta_l^{b^s_1}(1,2) \text{ etc.}\).
Now, the set $H$ contains $\alpha_i(x,y),\;i\geq 2$ and $\beta_j(x,y)$ for {\em all} $x\neq y$ and all $E_i$. Hence we may replace $\alpha_i(1,2)$ with $\alpha_(s,2)$, $\beta_j(1,2)$ with $\beta_j(s,2)$ and $E_1$ with $E_s$. This substitution will replace all $E_1\text{ and }D_1$ by $E_s$ and $D_s$ respectively. Now we may apply $E_s,D_s$ and $E_2$ in appropriate order to obtain $D_2$ in $H$. We next consider $\inv{\alpha_1}(2,1)$ and arguing as above we conclude that $D_1\in H$ and that the semigroup generated by $H$ is a subgroup.  
\end{proof}

We note that in case of insecure protocols the subgroup generated by $H$ is the full group generated by the encryption operators $\set{E_1,E_2,E_s}$ of the three parties concerned : the initiator, the intended receiver and the adversary. The theorem gives an abstract algebraic characterization of security. For practical purposes we would want a syntactic characterization in terms of the strings of operators. For this we start with a definition. 
\begin{Def}
Let $S=E_{j_1}^{i_1} E_{j_2}^{i_2} \cdots E_{j_k}^{i_k} $ be a string with $i_1, \dotsc , i_k$ integers and $j_1, \dotsc, j_k\in \set{1,\dotsc,n}$ in reduced form. For an integer $r$ in the set $\set{1,\dotsc,n}$ 
define the $r$-index of $S$ to be 
the sequence of integers $(r(1), r(2), \dotsc, r(m))$ which appear as {\tt nonzero} exponents of $E_r$ in $S$. 
We say that the $r$-index of $S$ is negative if the largest integer in the sequence is negative. 
\end{Def}

If the $r$ index of a string $S$ is negative then all the exponents of $E_r$ (there must be at least one) are negative. That is, only $D_r$ appears in $S$. Such strings are unbalanced as per \cite{DY}. Let us also say that $r-index$ is zero if no powers of $E_r$ appears in the string. Now we can state the second characterization of insecure protocols. 
\begin{thm} \label{thm:algo_secure1}
Let $\cali{P}$ a two-party cascade protocol. Assume that the legitimate parties have names $1$ and $2$ and the former initiates the conversation. Then $\cali{P}$ is insecure if and only if one of the following holds: 
\be
\item
The 2-index of $\alpha_1(1,2)$ is zero and the 1-index of $\alpha_1(1,2)$ is zero or negative. 
\item
There exists some $\alpha_i,\; i \geq 2$ whose  1-index is negative. 
 \item
There exists some $\beta_i,\; i \geq 1$ whose  2-index is negative.
\ee
\end{thm}
\begin{proof} {\em Sufficiency}. If the first condition above is satisfied then it is easy to see that the first condition in Theorem \ref{thm:alg_secure1} holds. Suppose now that the second or the third condition holds. We can use arguments similar to those in the previous theorem to show that $E_1$ and $E_2$ are in $S$ the semigroup generated by $H=\Gamma_1(s)\union \Gamma_2(x,y) \union \Gamma_3(x,y),\; x,y \in \set{1,2,s}$. 

\noindent
{\em Necessity}. The proof of necessity is rather long. We only outline the steps. Suppose $\cali{P}$ is insecure. From Theorem \ref{thm:alg_secure1} we infer that either the first condition holds or $S$ is a subgroup. If the first condition holds then clearly the 2-index of $\alpha_1(1,2)$ is zero and the 1-index of $\alpha_1(1,2)$ must be zero or negative. We may thus assume that $S$ is a subgroup. Then $E_1^{-1} \in S$. Write  $E_1^{-1}$ is a product of $\alpha_i(x,y),\; i \geq 2 $, $\beta_i(x,y),\; i \geq 1 $ and the $E_i$'s. We use induction on the length $l$ of such product. The case $l=1$ is clear. Let $l=k$. That is, $E_1^{-1}= \gamma\Phi$ where $\gamma \in H$ and $\Phi$ is in $S$. By assumption, none of the factors in $\Phi$ have negative $r$-index for $r\in \{1,2,s\}$. Now $\alpha_k(i,j)$ (resp.\ $\beta_k(i,j)$) cannot have negative $j$ (resp. $i$) index. Next show that if $\gamma_1,\gamma_2\in H$ have nonnegative $r$-index ($r=1,2$) then their product $\gamma_1\gamma_2$ also has nonnegative $r$-index. This is straightforward but lengthy. By assumption each of the generators of $S$ have nonnegative $r-$ index $r\in \{1,2\}$. Hence, as $\gamma$ and $\Phi$ have positive $r$-index for $r=1,2$ and so does $\gamma\Phi$, a contradiction. 
\end{proof}
The theorem yields the following corollary in some concrete realization of the cryptosystem. We recall that there may extra relations among the generators in any such realization. Let these relations be given by the set $R\subset F(\cali{E})$ where we put any $x\in R$ equal to $\epsilon$. Two strings in $F(\cali{E})$ are equivalent if they can be reduced to each other by insertion or deletion of elements from $R$. Then we have
\begin{cor}
A concrete realization of a two-party protocol is insecure if and only if each string in the equivalence classes of $\alpha_i,\; i>1$ and $\beta_j,\; j\geq 1$ has nonnegative 1 and 2 index. 
\end{cor}
\subsection{Algebraic characterization of security of general protocols}
In this section we will consider protocols with nonces (e.g.\ name-stamp). In the cascade protocols the structure of the plain text message itself played no role in the protocol. A name-stamp protocol uses the structure of the message to improve security. We use the notation as above. Now each user has more operations available. We have first the operation of nonce $A_x$ for user $x$: $A_x(M)= Mx$. We also have the partial inverse $\delta_{x}$, the deletion operator, that is, $\delta_xA_x(M)=M$. The problem is that it only makes sense to apply $\delta_x$ immediately after $A_x$ (after reduction in $E_i$s and $D_i$s). In fact, in \cite{DY} and other treatments \cite{DEK,EG} the application of $\delta_x$ is undefined in all other cases. However, for the algebraic structures we require that all products be well-defined.  Let $\cali{O}_x= \{ E_y,D_x, A_y, \delta_y| y \text{ any user }\}$ be the set of operators available to user $x$. Let $A$ be the set of operators $A_x$ and $\Delta$, the set of $\delta_x$s. We postulate the following relations:.\(E_xD_x  = D_xE_x= \epsilon \)
and \(\delta_x A_x = \epsilon \). Note that in this case we no longer have group since $A_x\delta_x\neq \epsilon$. 
\begin{Def}
A two-party name-stamp protocol is given by the following sequences of strings:
\(
\overline{\alpha}_i(x,y) \in (\{E_x, E_y, D_x\}\union A \union \Delta)^*, \quad \overline{\beta}_i(x,y) \in (\{E_x, E_y, D_y\}\union A \union \Delta)^*  \\
\) 
\end{Def}
We will assume that the protocol is well-defined, that is, there are no illegal operations of $\delta_x$. 
Let $\cali{O} = \union_x \cali{O}_x$ be the set of operators of all users. Let $G_O$ be the {\em free monoid} generated by $\cali{O}$. We are identifying $\inv{E_x}$ with $D_x$. We define \(\overline{N}_0(x,y)=\epsilon, \overline{N}_1(x,y) = \overline{\alpha}_1(x,y),\overline{N}_2(x,y) = \beta_1(x,y)\overline{\alpha}_1(x,y), \dotsc, \overline{N}_{2j-1}(x,y) = \overline{\alpha}_j(x,y)\overline{N}_{2j-2}(x,y)\) and \(\overline{N}_{2j}(x,y)= \beta_{j}(x,y)\overline{N}_{2j-1}(x,y)\) as before. We define a protocol to be {\em insecure} if there is a string $\gamma \in ({\Gamma'_1} \union {\Gamma'_2}\union {\Gamma'_3})^*$ such that $\gamma N_i(x,y)=\epsilon$ for some $i\geq 1$ where
\[
\begin{split}
\Gamma'_1(s) & = \{E_x ,E_s,D_s, A_x, \delta_x |\;x\in \{a,b,s\}\} \\
\Gamma'_2 & = \{\alpha_i(x,y)| x,y\in \{a,b,s\} \text{ and } i\geq 2 \} \\
\Gamma'_2 & = \{\beta_i(x,y)| x,y\in \{a,b,s\} \text{ and } i\geq 1 \} \\
\end{split}
\]
The motivation for the above definition of insecurity is similar to the case of cascade protocols. Excluding the trivial case (when the initiator $a$ sends the first string without encryption!) we state the algebraic characterization of security of these general protocols. 
\begin{thm} \label{thm:CriteName}
A name-stamp protocol is insecure if and only if $({\Gamma'_1} \union {\Gamma'_2}\union {\Gamma'_3})^*$ contains the  subgroup of $G_0$ freely generated by $\{E_a,E_b ,E_s\}$. 
\end{thm}
\begin{proof}({\em Sketch})
We observe first that, as in Theorem \ref{thm:alg_secure1} the condition for insecurity is equivalent to requiring 
that the string $\alpha_1(a,b)$ has an inverse. Clearly, the condition is sufficient since we can generate $D_a= \inv{E_a} \text{ and } D_b= \inv{E_b}$ and hence the inverse of any string. 

The necessity of the condition can proved using arguments similar to Theorem \ref{thm:alg_secure1}. We write $\inv{\alpha_1(a,b)}$ as a product of elements from ${\Gamma'_1} \union {\Gamma'_2}\union {\Gamma'_3}$. Then by appropriate changes $a\rightarrow s$ or $b\rightarrow s$ we can obtain $D_a$ and $D_b$. 
\end{proof}
The theorem implies, in particular, that the empty string $\epsilon$ is in $({\Gamma'_1} \union {\Gamma'_2}\union {\Gamma'_3})^+$, where for any set of strings $S$, $S^+=S^*-\set{\epsilon}$. The security of two-party ping-pong protocol is therefore equivalent to a decision problem for a regular language: is the empty string a member of the  language. For our case the problem is tractable. It is fairly straightforward to write an algorithm for the decision problem for the language $({\Gamma'_1} \union {\Gamma'_2}\union {\Gamma'_3})^+$ whose time complexity is bounded by polynomial in the length of the protocol. An efficient algorithm is given in \cite{DEK}. 

Let us consider some special protocols. 
\begin{propn}
Let a protocol $\prot$ be given by the following strings. $\alpha_1,\beta_1=\gamma_1\inv{\alpha_1},\alpha_2=\mu_2\inv{\gamma_1}, \beta_2=\gamma_2\inv{\mu_2}\cdots $ such that $\alpha_1,\gamma_i\text{ and } \mu_i$ have nonnegative 1 and 2 index, are not empty, do not contain any $
\delta_x$ and have their left-most symbol appropriate name-stamp $A_x$. Here $\inv{\sigma}$ denotes the left inverse of $\sigma$. Then $\prot$ is secure.  
\end{propn}
\begin{proof}({\em Sketch})
Suppose $\prot$ is insecure. Then there exist $v_1,v_2,\dotsc,v_k \in ({\Gamma'_1} \union {\Gamma'_2}\union {\Gamma'_3})$ such that $D_1= v_1\cdots v_k$. Then one of the $v_i$'s must be some $\alpha_i=\mu_i\inv{\gamma_{i-1}}$. But the right-most symbol of $\inv{\gamma_{i-1}}$ is a $\delta_x$. Hence, it must cancel. In fact, all the inverses must cancel. We are left with strings $\gamma_i$'s and $\mu_j$'s. But these have nonnegative 1-index and from the previous section one cannot obtain $D_1$ with these generators.  
\end{proof}
We can similarly show that if in some protocol $\prot$ we have some $\alpha_i(1,2),\;i>1$($\beta_j(1,2),\;j\geq 1$) such that the substrings on the left and right of the left-most $\delta_2$($\delta_1$) have negative 1-index(2-index) then the protocol is insecure. We only have to consider $\alpha_i(1,s)$ and cancel appropriate symbols using $D_s,A_s,\text{ and } E_1$. 
\subsection{Examples, Extensions, and Concrete Realizations} 
Consider now a simplified variant of Needham-Schroeder authentication protocol \cite{Lowe}. We have $\alpha_1(1,2)=E_2A_1,\beta_(1,2)= E_1\delta_2D_2\text{ and }\alpha_2(1,2)=E_2D_1$. In detail, user 1 stamps its nonce and sends the string to 2 using the latter's public key encryption $E_2$. User 2 then decrypts the message and sends it back to 1 using its public encryption and 1 decrypts the message and sends it to 2 after encryption. We see at once that the protocol is insecure because $\alpha_2(1,2)=E_2D_1$ has negative 1-index. We observe that the reason it is insecure is because there is no nonce in stage 2. Hence, if we modify the protocol \cite{Lowe} with $\alpha_1(1,2)=E_2A_1,\beta'_(1,2)= E_1A_2\delta_1D_2\text{ and }\alpha'_2(1,2)=E_2\delta_2 D_1$ from the above proposition it follows that the protocol is secure. On the other hand, following protocol \cite{DY} is insecure:  $\alpha_1(1,2)= E_2A_1E_2, \beta_1(1,2)= E_1A_2D_2\delta_1D_2$, since in $\beta_1(1,2)$ the substrings to left and right of $\delta_1$ have negative 2-index. We therefore observe that with the use of above propositions we can eliminate large classes of protocols as insecure. Although we do not have a necessary and sufficient criterion for security (as in the case of ping-pong protocols) we can write efficient algorithms to verify security. These are essentially rewriting algorithms  in groups \cite{Sims}. 

We investigated the algebraic structures arising out of protocols based on public or asymmetric key cryptosystem. Can we extend this to private or symmetric key cryptography. In case of, two party protocols the answer is yes. If users 1 and 2 share a private key then we set $E_1=E_2$ and remove $E_1$ from adversary's set of operations $\Gamma$ (see the previous section). The security of the protocols is defined as above. 

A (concrete) {\em realization} of an abstract protocol is a map $\phi: G_{O} \rightarrow G$ which is monoid homomorphism. Here $G_O$ is the free monoid on the set $O$ of operations available to all users and $G$ is some monoid. Any map from $O$ to $G$ can be uniquely extended to a homomorphism $\phi: G_{O} \rightarrow G$. In general, $G$ will satisfy some extra relations. For example, if $G$ is finite then for any $x\in G,\; x^{|G|}=e$ Then, the security criteria of Theorem \ref{thm:CriteName} is inadequate since any subset of $G$ will generate nontrivial subgroups. An example is the cyclic subgroup $\{ E_1,E_1^2, \dotsc, E_1^{|G|}=e\}$. Hence, we must modify the security condition. Our proposal is to require the relevant groups be only {\em pseudo-free} \cite {Rivest,Micc} instead of free. Informally, a group $G$ is pseudo-free if any polynomial time probabilistic algorithm designed to find relations in $G$ that are not satisfied in a free group will succeed with only negligible probability.  Let $\cali{P}$ be a two-party protocol and let $\Gamma$ be the set of operators (in reduced form) available to an adversary as in the preceding sections. Then $\phi(\Gamma)$ may contain non-trivial groups. Suppose all these groups are pseudo-free. Then any special relations that the adversary may try to exploit can only be found with negligible probability by any feasible algorithm. We note that the security of a protocol may be compromised in two ways. First, the adversary may break the cryptosystem itself, for example, by finding an efficient algorithm to factorize integers in RSA-based cryptosystem. The second way is to exploit some weakness in the protocol itself as in the Needham-Schroeder protocol. Both cases are covered by the following definition. 

\begin{Def}
A protocol $\cali{P}$ is insecure if and only if one of the following holds.  
\be
\item
In the free group $\Gamma$, the set of operations available to the adversary generate a nontrivial subgroup. 
\item
The maximal subgroup contained in the monoid generated by $\Gamma_1' \bigcup \Gamma_2'\bigcup \Gamma_3'$ in a family of concrete realizations of the encryption and decryption operators is {\em not} pseudo-free. 
\ee
\end{Def}
If the basic public key cryptosystem is RSA then in general the encryption operators $E_i$ are based on different moduli and the messages may have to be split into blocks of appropriate size before each encryption. The operators $E_i$ are quite complicated and form a non-abelian group. In the ElGamal encryption scheme \cite{ElGamal} the encryption operator is a map $E_a:Z^*_p\rightarrow Z^*_p$ where $E_a(m)=mg^{x_ak}$ . All operations are modulo $p$, $g$ is a primitive generator of $Z^*_p$, $g\text{ and} g^{x_a}$ are publicly known. The number $k$ is randomly chosen by $b$ and $g^{k}$ is publicly known. The adversary does not know $k$ or $x_a$ and hence $E_a$. This is similar to the case of private key cryptosystem since we have to remove $E_a$ from the set of operations available to adversary. If all users use the same $p$ the group is abelian. However, if they choose different primes the messages have to be block  and the resulting realization of the DY group is non-abelian in general. 

\section{Discussion}
In this work we presented an algebraic characterization of security of public key protocols. We may question the advantages of the algebraic characterization. First, there are theoretical advantages. We have at our disposal powerful techniques of group theory. To prove some fact in the setting of free groups we can define a homomorphism from the free group to  another (not necessarily free) group which has a simpler structure. For example, in Theorem \ref{thm:algo_secure1} we defined the notion of $r$-index and stated that it is positive for the product of two strings whose $r$-index is positive. The proof is given by induction and a tedious case by case consideration on the structure of the two strings. It is possible to give a group theoretic proof by defining a homomorphism to another group via some defining relations. Secondly, there are practical advantages too. Sometimes, often computations and rewriting in groups is simpler and we have at our disposal several computational tools \cite{Sims}. 

This work is an attempt to give a new, algebraic perspective on security and there is still a lot of ground to be covered. Can we extend the formal algebraic characterization to other protocols? An essential requirement for group structure is that all the operations be invertible. For example, could also include operations like pairing. We then just have the structure of a monoid, as in the case of name-stamp protocols and we have seen that these can be dealt with in an algebraic setting. We aim to deal with these issues in the future. 


\begin{thebibliography}{MW04b}

\bibitem[AR02]{AbadiRog}
M.~Abadi and P.~Rogaway.
\newblock Reconciling two views of cryptography.
\newblock {\em J. of Cryptology}, 15(2):103--127, 2002.

\bibitem[BAN89]{Ban}
M.~Burrows, M.~Abadi, and R.~Needham.
\newblock A logic of authentication.
\newblock In {\em Proc. Royal. Soc. Lond. A.}, pages 426:233--271, 1989.

\bibitem[DEK82]{DEK}
D.~Dolev, S.~Even, and R.~M. Karp.
\newblock On the security of ping-pong protocols.
\newblock {\em Inform. and Control}, 55:57--68, 1982.

\bibitem[DY82]{DY}
D.~Dolev and A.~C. Yao.
\newblock On the security of public key protocols.
\newblock {\em IEEE Trans. Inform. Theory}, IT-30(2):198--206, 1982.

\bibitem[EG83]{EG}
S.~Even and O.~Goldreich.
\newblock On the security of multiparty ping-pong protocols.
\newblock Research Report TR-04-02, Comp Sc. Dept., Tecnicion, Haifa, 1983.

\bibitem[ElG85]{ElGamal}
T.~ElGamal.
\newblock A public key encryption and signature scheme based on discreet
  logarithm.
\newblock In {\em Proc. of Crypto 84, LNCS 196}, pages 10--18. Springer, 1985.

\bibitem[Gol01]{Goldreich}
O.~Goldreich.
\newblock {\em Foundations of cryptography: basic tools}.
\newblock Cambridge University Press, 2001.

\bibitem[Low96]{Lowe}
G.~Lowe.
\newblock Breaking and fixing the needham-schroeder public-key protocol using
  fdr.
\newblock In {\em Lect. Notes. Comp. Sc., 1055}, pages 147--166. Springer,
  1996.

\bibitem[Mic05]{Micc}
D.~Micciancio.
\newblock The rsa group is pseudo-free.
\newblock In {\em Proc. of Eurocrypt 2005, LNCS 3494}, pages 387--403.
  Springer, 2005.

\bibitem[MKS76]{MKS}
W.~Magnus, A.~Karras, and D.~Solitar.
\newblock {\em Combinatorial group theory}.
\newblock Dover, 1976.

\bibitem[MW04a]{WarMicc1}
D.~Micciancio and B.~Warinschi.
\newblock Completeness theorems for abadi-rogaway logic of encrypted
  expressions.
\newblock {\em J. of Comp. Security}, 15:99--121, 2004.

\bibitem[MW04b]{WarMicc2}
D.~Micciancio and B.~Warinschi.
\newblock Soundness of formal encryption in the presence of active adversaries.
\newblock In {\em Proc. of TCC (Theory of Cryptography Conferebce) 2004, LNCS
  2951}, pages 133--151. Springer, 2004.

\bibitem[NS78]{NS}
R.~M. Needham and M.~D. Schroeder.
\newblock Using encryption for authentication in large network computers.
\newblock {\em Comm. of the ACM}, 21(2):993--999, 1978.

\bibitem[Riv04]{Rivest}
R.~L. Rivest.
\newblock The notion of pseudo-free groups.
\newblock In {\em Proc. of TCC 2004, LNCS 2951}, pages 505--521. Springer,
  2004.

\bibitem[Rot95]{Rotman}
Joseph~J. Rotman.
\newblock {\em An introduction to the theory of groups}.
\newblock Springer-Verlag, 1995.

\bibitem[Sim94]{Sims}
C.~C. Sims.
\newblock {\em Computations with finitely presented groups}.
\newblock Cambridge University Press, 1994.

\end{thebibliography}
\end{document}